\documentstyle [12pt,epsfig]{article}
\title{Natural Quark Mass Patterns}
\author{K. Wang 
\\ Dept. of Physics, University of California, Los Angeles 
\\ Los Angeles, California 90024}
\date{}
\def\gsim{\mathrel{\raise.3ex\hbox{$>$\kern-.75em\lower1ex\hbox{$\sim$}}}}
\def\lsim{\mathrel{\raise.3ex\hbox{$<$\kern-.75em\lower1ex\hbox{$\sim$}}}}

\begin{document}
\maketitle

\begin{abstract}
We incorporate the idea of natural mass matrices into the construction of 
phenomenologically viable quark mass matrix patterns. The general texture 
pattern for natural Hermitian mass matrices is obtained and several 
applications of this result are made.  
\end{abstract}

\vspace{4.5cm}
\begin{flushleft}
$\quad$UCLA/96/TEP/14

$\quad$April 1996
\end{flushleft}
\pagebreak

Recently, we proposed the idea of natural mass matrices~\cite{MM}, an 
organizing principle useful in the construction of phenomenologically viable
grand unification theory (GUT) scale quark mass matrix patterns. In this 
note, we present a detailed implementation and discuss certain
applications of this result, among which is the construction of some
supersymmetric (SUSY) GUT mass matrix patterns. We begin with a brief summary
of the low energy data (LED) which we use as inputs and a discussion of the
evolution of these parameters in the minimal supersymmetric standard model 
(MSSM).  This is followed by the introduction of a convenient parametrization 
of Hermitian mass matrices which, along with the ``naturalness'' 
requirement~\cite{MM}, allows us to derive the general texture pattern for 
natural Hermitian quark mass matrices. The usefulness of this result is then 
demonstrated through several examples. Specifically, using the expression for 
this general pattern, we conduct an efficient viability check on a known quark 
mass pattern, perform an exercise of finding mass patterns with most 
texture-zeros and finally, we construct some simple, generic mass patterns 
which may be useful as templates for contemplating ``predictive''  quark mass 
Ansatze. 
\bigskip

\begin{flushleft}
{\bf 1. LED Inputs and Their Evolution in MSSM}
\end{flushleft}

In our bottom-up approach of constructing quark mass matrices, we use as 
inputs the quark mass ratios evaluated at $m_{t} \simeq 175$ (GeV) and values 
of CKM matrix elements in the standard Wolfenstein 
parametrization~\footnote{See Ref.~\cite{MM} for the sources of the numbers 
summarized in Table~\ref{LED}.}:

\scriptsize
\begin{table}[h!]
\begin{center}
\begin{tabular}{|c|c|c|c|}
\hline & \\
Quark mass ratios & $\begin{array}{ll}
m_{u}/m_{t}=\xi_{ut}\lambda^{7}\phantom{\lambda} & \xi_{ut} = 0.49 \pm 0.15 \\ 
m_{c}/m_{t}=\xi_{ct}\lambda^{4} & \xi_{ct} = 1.46 \pm 0.13 \\
m_{d}/m_{b}=\xi_{db}\lambda^{4} & \xi_{db} = 0.58 \pm 0.18 \\
m_{s}/m_{b}=\xi_{sb}\lambda^{2} & \xi_{sb} = 0.55 \pm 0.18 \phantom{\delta 
\simeq  [45^{0}, \; 158^{0}]}
\end{array}$ \\ & \\ \hline & \\
CKM parameters & $\begin{array}{ll}
V_{us} = \lambda + O(\lambda^{7}) & \lambda = 0.221 \pm 0.002  \\
V_{cb} = A\lambda^{2} + O(\lambda^{4}) & A = 0.78 \pm 0.05 \\
V_{ub} =  A \sigma \lambda^{3} e^{-i\delta} & \sigma = 0.36 \pm 0.09,\; \delta 
\simeq  [45^{0}, \; 158^{0}] 
\end{array}$ \\ & \\ 
\hline
\end{tabular} 
\caption{LED inputs used in quark mass pattern construction.} \label{LED}
\end{center}
\end{table}
\normalsize

Notice, in particular, the different degrees of experimental uncertainties 
associated with the LED. Roughly speaking, $\Delta \lambda$ is about $1\%$, 
$\Delta A, \Delta \xi_{ct}$ are slightly below $O(10\%)$ while $\Delta \sigma, 
\Delta \xi_{ut}, \Delta \xi_{db}, \Delta \xi_{sb}$ are of $O(30\%)$ and, 
$\delta$ is only loosely bounded.

Additionally, one can impose the existing constraints on the relative sizes of 
the light quark masses from current algebra analyses~\cite{KMLeut} 
\begin{equation}
(\frac{m_{u}}{m_{d}})^{2}+\frac{1}{Q^{2}}(\frac{m_{s}}{m_{d}})^{2} = 1 \;, 
\;\;\; \mbox{with} \;\; Q = 24 \pm 1.6 \; . \label{light_q}
\end{equation} According to a recent study~\cite{Leut}, the value of $Q$ in 
the above equation is likely to be somewhat smaller ($Q = 22.7 \pm 0.8$); and,
 the range of values for the quark mass ratios may be even further narrowed 
down to: $m_{u}/m_{d}=0.553 \pm 0.043$ and $m_{s}/m_{d}=18.9 \pm 0.8$. In 
terms of the mass ratio parameter $\xi$'s of Table~\ref{LED}, the latter 
translates to \begin{equation}
\xi_{db}/\xi_{sb}  =  1.09 \pm 0.04 \;\; . \label{mds} 
\end{equation}

Typically, one wishes to construct mass patterns at some high energy scales 
as, for instance, one does when building certain GUT models. To do so using 
the LED inputs of Table~\ref{LED}, one must also take into account the 
evolution of these parameters. Here, as an example, we consider the scaling 
behavior of the LED parameters in the MSSM which actually has a rather simple 
description, provided that the underlying mass matrices are ``natural'' 
~\cite{MM}.~\footnote{Assuming quark and lepton mass matrices are ``natural'', 
the corresponding Yukawa matrices then exhibit a certain definite hierarchy. In 
particular, the [3,3] matrix elements are much greater than the rest -- a fact 
gainfully exploited in the simplification of solutions to the one-loop 
renormalization group equations (RGE's) for the Yukawa matrices~\cite{SBOP}.} 
Denoting the [3,3] matrix elements of the u-type and d-type Yukawa matrices as 
$\lambda_{u}$ and $\lambda_{d}$ respectively, one has~\footnote{For conciseness, 
unless otherwise specified, values of the parameters in the expressions below 
are taken to be those evaluated at $m_{t}$.}  \begin{eqnarray}
\xi_{ct}(m_{G})/\xi_{ct} & \simeq & \xi_{ut}(m_{G})/\xi_{ut} \simeq r_{u} \;,\; 
\label{ru} \\
\xi_{sb}(m_{G})/\xi_{sb} & \simeq & \xi_{db}(m_{G})/\xi_{db} \simeq r_{d} 
\label{rd} \;,\; \\
\lambda(m_{G})/ \lambda  & \simeq & \sigma(m_{G})/ \sigma  \simeq 
1 \;,\; \\
A(m_{G})/A & \simeq & r \label{rr}
\end{eqnarray} where the scaling parameters are defined by \begin{eqnarray}
r_{u} & = & e^{-\frac{1}{16\pi^{2}} \int_{\ln m_{t}}^{\ln m_{G}}\{3\lambda_{u}^{2}(\mu) +\lambda_{d}^{2}(\mu)\} \, d\ln \mu} \;, 
\nonumber \\
r_{d} & = & e^{-\frac{1}{16\pi^{2}} \int_{\ln m_{t}}^{\ln m_{G}} 
\{\lambda_{u}^{2}(\mu) + 3\lambda_{d}^{2}(\mu) \} \, d\ln \mu} \;, \nonumber \\
r & = & e^{-\frac{1}{16\pi^{2}} \int_{\ln m_{t}}^{\ln m_{G}} 
\{\lambda_{u}^{2}(\mu) +\lambda_{d}^{2}(\mu)\} \, d\ln \mu} \;. \label{r}
\end{eqnarray} The $\lambda(\mu)$'s in these expressions are furthermore 
determined from the RGE's \begin{eqnarray}
\frac{d\lambda_{u}}{d\ln\mu} & \simeq & \frac{1}{(4\pi)^{2}} \left\{ 
6\lambda_{u}^{2}+\lambda_{d}^{2}-c_{i}g^{2}_{i}\right\} \lambda_{u}  \; , 
\nonumber \\
\frac{d\lambda_{d}}{d\ln\mu} & \simeq & \frac{1}{(4\pi)^{2}} \left\{ 
6\lambda_{d}^{2}+\lambda_{u}^{2}+\lambda_{e}^{2}-c'_{i}g^{2}_{i} \right\} 
\lambda_{d} \; , \nonumber \\
\frac{d\lambda_{e}}{d\ln\mu} & \simeq & \frac{1}{(4\pi)^{2}} \left\{ 
4\lambda_{e}^{2}+3\lambda_{d}^{2}-c''_{i}g^{2}_{i} \right\} \lambda_{e} \;, 
\nonumber \\
\frac{dg_{i}}{d\ln\mu} & \simeq & \frac{1}{(4\pi)^{2}} \, b_{i}g_{i}^{3} 
\quad\quad (i=1,2,3) \;.  
\label{RGEs}
\end{eqnarray} Here, $\lambda_{e}$ denotes the [3,3] matrix element of the 
lepton Yukawa matrix and, $c_{i}=(13/15,\,3,\,16/3)$, $c'_{i}=(7/15,\,3,\,16/3)$, 
$c''_{i}=(9/5,\,3,\,0)$, $b_{i}=(33/5,\,1,\,-3)$. 
The scale $m_{G} \simeq 
10^{16}$ (GeV) is the unification point of the three gauge couplings which we 
shall take to be $\alpha_{i}=(0.017,\,0.033,\,0.100)$ (with $\alpha_{i} \equiv 
g_{i}^{2}/4\pi$) following Ref.~\cite{DHR}. 

Further simplification is possible if one assumes that $\tan \beta \ll 
O(m_{t}/m_{b})$, in which case the contributions of the $\lambda_{d}$ and 
$\lambda_{e}$ terms in Eq.~(\ref{r}) can be largely neglected and as a result, 
$r_{d} \simeq r$ and $r_{u} \simeq r^{3}$. In the same limit, the evolution of 
$\lambda_{u}$ and $\lambda_{d}$ is given by \begin{eqnarray*} 
\lambda_{u}(\mu)/\lambda_{u} & \simeq & \{\eta(\mu)\}^{1/2} \, \{ 1-(3/4\pi^{2}) \, 
\lambda_{u}^{2} \, I(\mu)\}^{-1/2}  \;, \\
\lambda_{d}(\mu)/\lambda_{d} & \simeq & \{\eta'(\mu)\}^{1/2} \,\{ 
\lambda_{u}(\mu)/\lambda_{u}\}^{1/6} \, \{\eta(\mu)\}^{-1/12} 
\end{eqnarray*} where \[ \eta(\mu) \equiv  
\prod^{i}\{\alpha_{i}/\alpha_{i}(\mu)\}^{c_{i}/b_{i}} \;\;,\;\; \eta'(\mu) 
\equiv \prod \{\alpha_{i}/\alpha_{i}(\mu)\}^{c'_{i}/b_{i}} \; , \] and \[ I(\mu) 
\equiv \int_{\ln m_{t}}^{\ln \mu} \eta(\mu) \, d \ln \mu \;\;.\] Expressed in 
terms of the above parameters and functions, one has \begin{equation}
r \simeq \left\{ \lambda_{u}(m_{G})/\lambda_{u} \right\}^{-1/6} 
\{\eta(m_{G})\}^{1/12} \;. \label{r_approx}
\end{equation} From these results one sees  that $r \simeq O(1)$ except when 
$\lambda_{u}$ approaches a small region defined by $\lambda_{u} \simeq 
2\pi/\sqrt{3 I(m_{G})}$ where $r$ rapidly drops to zero.  

For large $\tan \beta$'s, the analysis becomes more involved and  one has to 
rely upon numerical methods for solving Eq.~(\ref{RGEs}) and evaluating the 
$r$'s in Eq.~(\ref{r}). Interestingly, the values of the $r$'s do not deviate 
much from being of $O(1)$ unless $\tan \beta $ reaches near the value of 
$m_{t}/m_{b}$ where they begin to drastically decrease again. For a qualitative 
understanding of this observation, we solve Eq.~(\ref{RGEs}) with the assumption 
$\lambda_{u}=\lambda_{d}$ (corresponding to $\tan \beta = m_{t}/m_{b}$)  while 
momentarily ignoring contributions from the leptonic sector.  In this limit, we 
find \[ \lambda_{u}(\mu) \simeq \lambda_{d}(\mu) \propto \{ 
1-(3.5/4\pi^{2})I(\mu) \}^{-1/2}  \] which yields approximately 
$\lambda_{u}(m_{G}), \; \lambda_{d}(m_{G}) \rightarrow \infty$. Refering 
moreover to the results in Eqs.~(\ref{r}) and (\ref{r_approx}), we have then 
$r_{u} \simeq r_{d} \simeq r^{2}$ with $r \rightarrow 0$.  

\begin{figure}
~\epsfig{file=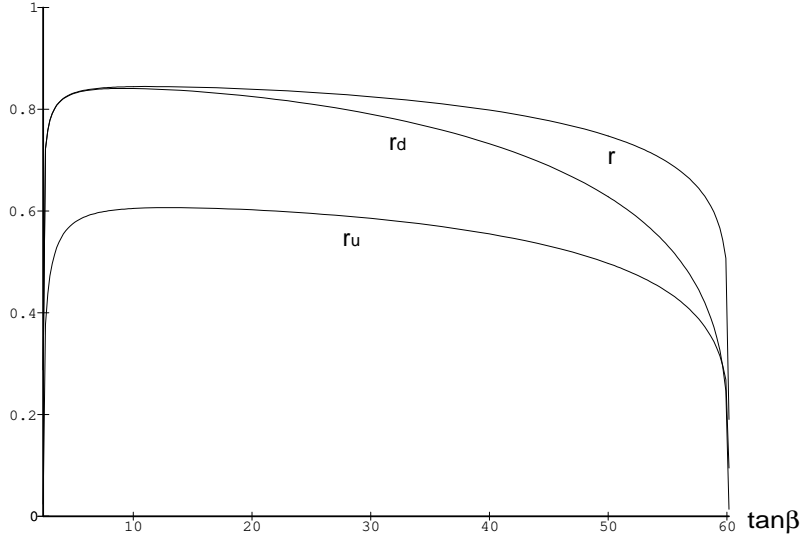,width=5cm}
\caption{Scaling parameters as functions of $\tan\beta$} \label{rrr}
\end{figure}

For easy reference, we include in Fig.~\ref{rrr} a plot based on numerical 
solutions of Eqs.~(\ref{r}) and (\ref{RGEs}) subject to the boundary condition 
(at $m_{t}$)~\footnote{More detailed results of some related calculations based 
on two-loop RGE's can be found in Ref.~\cite{Barger}.} 
\[m_{t}=\frac{v}{\sqrt{2}}\lambda_{u}\sin\beta \;,\; 
m_{b}=\frac{v}{\sqrt{2}}\lambda_{d}\cos\beta \;,\; 
m_{\tau}=\frac{v}{\sqrt{2}}\lambda_{e}\cos\beta\] with $v \simeq 246.2$ (GeV), 
$m_{t} \simeq 175$ (GeV), $m_{b} \simeq 2.78$ (GeV) and $m_{\tau} \simeq 1.76$ 
(GeV) \footnote{Notice the end regions of the plot below depend rather 
sensitively on the exact values of the numbers taken.}.

\bigskip
 
\begin{flushleft}
{\bf 2. The General Texture Pattern of Natural Hermitian Quark Mass Matrices}
\end{flushleft}

\begin{flushleft}
(i) {\it A Parametrization of Hermitian Quark Mass Matrices}
\end{flushleft}

Given the (scaled) diagonal quark mass matrices~\cite{MM} \small \begin{equation}
\tilde{M}_{u}^{diag}(m_{t}) = \left( \begin{array}{ccc} 
 \xi_{ut}\lambda^{7} &  0      & 0 \\  
 0      &  \xi_{ct}\lambda^{4} & 0 \\  
 0      &  0      &  1
\end{array} \right) \;, \quad \tilde{M}_{d}^{diag}(m_{t}) = \left( \begin{array}{ccc} 
 \xi_{db}\lambda^{4} &  0      & 0 \\  
 0      &  \xi_{sb}\lambda^{2} & 0 \\  
 0      &  0      & 1  
\end{array} \right)
\end{equation} \normalsize and the CKM matrix (in the standard form)~\cite{PDG} \small
\begin{equation}
V = \left( \begin{array}{ccc} 
 c_{1}c_{3}   &  s_{1}c_{3}     &  s_{3}e^{-i\delta}  \\ 
 -s_{1}c_{2}-c_{1}s_{2}s_{3}e^{i\delta} &  c_{1}c_{2}-s_{1}s_{2}s_{3}e^{i\delta}     
&  s_{2}c_{3}  \\
 s_{1}s_{2}-c_{1}c_{2}s_{3}e^{i\delta} &  -c_{1}s_{2}-s_{1}c_{2}s_{3}e^{i\delta}     
&  c_{2}c_{3}  
 \end{array} \right) \;\; , \label{CKM}
\end{equation} \normalsize the most general Hermitian mass matrices can be 
constructed from \begin{eqnarray}
\tilde{M}_{u} & = & U \, \tilde{M}_{u}^{diag} \, U^{\dagger} \; , \label{Mu} \\
\tilde{M}_{d} & = & D \, \tilde{M}_{d}^{diag} \, D^{\dagger} \label{Md}
\end{eqnarray} in which the unitary matrices $U, \; D$ are subject to the 
constraint \begin{equation} 
U^{\dagger}D = \Phi_{u} \, V \, \Phi_{d} 
\end{equation} where $\Phi_{u,d}$ are some diagonal phase matrices. Furthermore,  
aside from a trivial quark-phase redefinition (which amounts to $U 
\leftrightarrow \Psi U$ and $D \leftrightarrow \Psi D$, $\Psi$ being some common 
phase matrix), one can always, for example, choose to parametrize the unitary 
matrices $U,\,D$ according to: \begin{equation}
\mbox{(i)}\quad\quad\quad\quad D \rightarrow D_{s} \quad \mbox{and then}\quad 
U^{\dagger} \rightarrow V \Phi D_{s}
\end{equation} or somewhat analogously,  \begin{equation} 
\mbox{(ii)}\quad\quad\quad\quad U^{\dagger} \rightarrow U^{\dagger}_{s} \quad 
\mbox{and then}\quad D \rightarrow U_{s} \Phi' V \; \label{CKM_constraint}
\end{equation} where $D_{s}$, $U^{\dagger}_{s}$ are the matrices $D$, 
$U^{\dagger}$ written in the standard form (of $V$) after necessary 
rephasing, and $\Phi,\,\Phi'$ are some diagonal phase matrices which can be 
arranged to have only two phases in each. 

In what follows, we shall adopt prescription (ii) since we have found it to be 
more convenient for constructing natural mass matrices~\footnote{Previously, in 
Ref.~\cite{MM}, we followed (i) to construct several mass pattern examples.}. 
Specifically, we let \begin{equation} \Phi' \equiv  \left( \begin{array}{ccc}
e^{i\phi}  & 0 & 0 \\
0  & e^{i\psi} & 0 \\
0  & 0 & 1
\end{array} \right)  \label{Phi} 
\end{equation} and write, in terms of some orthogonal rotation matrices ($C$'s) 
and some diagonal phase matrices ($\Delta$'s)~\footnote{See Ref.~\cite{MM} for 
the precise definitions of the matrices introduced below.} \[ V = C_{2} \, 
\Delta \, C_{3} \, \Delta^{\dagger} C_{1} \] and accordingly \[ U =  C_{1u} \, 
\Delta_{u} \, C_{3u} \, \Delta^{\dagger}_{u} C_{2u} \; . \] Defining three more 
orthogonal matrices $C_{id} \equiv C_{iu} \, C_{i} \;  (i=1,2,3)$ we have, by 
Eq.~(\ref{CKM_constraint}), \begin{eqnarray*} 
D & = &  \{C_{1d}\} \, \{ C^{\dagger}_{1} \, (\Delta_{u} \, C_{3d} \, 
\Delta^{\dagger}_{u}) \, C_{1} \} \nonumber \\
  & & \{ C^{\dagger}_{1} \, (\Delta_{u} \, C^{\dagger}_{3} \, 
\Delta^{\dagger}_{u}) \, ( C_{2d} \, C^{\dagger}_{2} \, \Phi' \, C_{2} ) \, 
(\Delta \, C_{3} \, \Delta^{\dagger}) \, C_{1} \} \; .
\end{eqnarray*}
\smallskip

\begin{flushleft}
(ii) {\it The Texture Pattern of Natural Mass Matrices}
\end{flushleft}

Following the procedure described in Ref.~\cite{MM}, we proceed to express $V$ 
in the Wolfenstein parametrization~\cite{WOLF} and likewise the matrices $C$'s 
as perturbative expansions in terms of the small parameter $\lambda$ 
(Table~\ref{LED}). Subsequently, we apply our ``naturalness'' 
criterion~\cite{MM} on the resulting quark mass matrices to arrange for natural 
mass patterns. Below we summarize our main result.  

In terms of the CKM matrix parameters ($\lambda, \; A, \; \Lambda\equiv\sigma 
A/\lambda, \; \delta$), the quark mass ratios ($\xi$'s) and the free phases 
($\phi, \; \psi$) introduced in Eq.~(\ref{Phi}), natural Hermitian quark mass 
matrices exhibit the following general texture pattern \begin{eqnarray}
\tilde{M}_{u} & = & \left( \begin{array}{ccc} 
 u_{11}\lambda^{7} &  u_{12}\lambda^{6}  &  u_{13}\lambda^{4}  \\   
 u^{*}_{12}\lambda^{6} &  u_{22}\lambda^{4}  &  u_{23}\lambda^{2} \\
 u^{*}_{13}\lambda^{4} &  u^{*}_{23}\lambda^{2}  &  u_{33} 
 \end{array} \right) \nonumber \\ \nonumber \\
\tilde{M}_{d} & = &  \left( \begin{array}{ccc} 
 d_{11}\lambda^{4} &  d_{12}\lambda^{3}  &  d_{13}\lambda^{4}  \\ 
 d^{*}_{12}\lambda^{3} &  d_{22}\lambda^{2}  &  d_{23}\lambda^{2} \\
 d^{*}_{13}\lambda^{4} &  d^{*}_{23}\lambda^{2}  &  d_{33} 
 \end{array} \right) \label{GP}
\end{eqnarray} with,~\footnote{In specifying the values of quark masses or mass 
ratios we conventionally quote these numbers as being positive. In the 
expressions below and throughout the presentation of our results however, quark 
masses $m_{q}$'s (and in general quark mass ratios $\xi$'s as well) can be 
chosen to have either positive or negative signs, depending upon the context of 
expressions they are in.} \begin{eqnarray}
u_{11} & = & \xi_{ut} + \{\alpha^{2}\xi_{ct}+|u_{13}|^{2}\} \lambda + O(\lambda^2) 
\;\; , \nonumber \\
u_{12} & = & \alpha \xi_{ct} + u_{13} u_{23} + O(\lambda^2) \;\; , \nonumber \\
u_{22} & = & \xi_{ct} + |u_{23}|^{2} + O(\lambda^2) \;\; , \nonumber \\
u_{33} & = & 1 + O(\lambda^{4}) \;\; ; \nonumber \\
d_{11} & = & \xi_{db} + |d_{12}|^{2}/\xi_{sb} +  O(\lambda^2) \;\; , \nonumber \\
d_{12} & = & \xi_{sb} \{e^{i(\phi-\psi)}+ \alpha \, \lambda\}  +  O(\lambda^2) \;\; , 
\nonumber \\
d_{13} & = & u_{13} + \alpha A e^{i\psi} + \Lambda e^{i(\phi-\delta)} - d_{12}d_{23} 
\, \lambda  +  O(\lambda^2) \;\; , \nonumber \\ 
d_{22} & = & \xi_{sb} +  O(\lambda^2) \;\; , \nonumber \\
d_{23} & = & u_{23} + A e^{i\psi}  +  O(\lambda^2) \;\; , \nonumber \\
d_{33} & = & 1 + O(\lambda^{4}) \;\; . \label{constraint_1}
\end{eqnarray} The remaining matrix element parameters in Eq.~(\ref{GP}) are not 
fixed, but are constrained by our requirement of ``naturalness'' to: \begin{equation}
|u_{13}|, \; |u_{23}|, \; \alpha \; \lsim O(1) \;\; . \label{constraint_2}
\end{equation} 

The quark mass matrices of Eq.~(\ref{GP}), although defined apparently at the 
scale of $m_{t}$, can nonetheless be implemented at any energy scale so long as 
one properly takes into account the renormalization group (RG) evolution of the 
quark mass ratios and the CKM parameters in Eq.~(\ref{constraint_1}). 
\bigskip

\begin{flushleft}
{\bf 3. Applications}
\end{flushleft}

\begin{flushleft}
(i) {\it Mass Pattern Viability Check}
\end{flushleft}

Given any natural mass pattern, once written in the form of Eq.~(\ref{GP}), one 
can examine its viability using Eqs.~(\ref{constraint_1}) and 
(\ref{constraint_2}). Specifically, the matrix elements of the pattern must, to 
a good approximation, obey the follwing constraints: \begin{eqnarray} 
&& u_{22} - |u_{23}|^{2} \simeq  \xi_{ct} \;,\nonumber \\
&& d_{22} \simeq \xi_{sb} \;,\nonumber \\
&& |d_{23} - u_{23}| \simeq A \;, \nonumber \\ 
&& |d_{12}/\xi_{sb} - (u_{12}-u_{13}u_{23})\lambda/\xi_{ct}|  \simeq  1 \;, 
\nonumber \\
&& |d_{13} - u_{13} - (d_{23}-u_{23})(u_{12}-u_{13}u_{23})/\xi_{ct} + 
d_{12}d_{23}\,\lambda| \simeq  \Lambda  \;, \nonumber \\
&& u_{11} - |u_{13}|^{2} \lambda - (|u_{12}-u_{13}u_{23}|^{2}/\xi_{ct})\, 
\lambda - \xi_{ut}  \simeq  0 \;, \nonumber \\
&& d_{11} - (\xi_{db} + |d_{12}|^{2}/\xi_{sb}) \simeq 0 \;, \nonumber \\  
&& \arg\{d_{13}-u_{13}-(d_{23}-u_{23})(u_{12}-u_{13}u_{23})/\xi_{ct} + 
d_{12}d_{23}\,\lambda\} \nonumber \\
&& - \arg\{d_{12}/\xi_{sb} - (u_{12}-u_{13}u_{23})\lambda/\xi_{ct}\} - 
\arg\{d_{23}-u_{23}\} \simeq - \delta \;. \label{check}
\end{eqnarray} 

As an illustrative example, we apply the results of Eq.~(\ref{check}) to the 
study of a mass pattern recently proposed based on the idea of a ``democratic'' 
symmetry~\cite{Fritz}. In this model, after some straightforward manipulations, 
the quark mass matrices take the form \small \[ M_{u} \simeq m_{t} \left( 
\begin{array}{ccc} 
 0 &  u  &  0  \\   
 u &  \frac{2}{9} \epsilon_{u}  & -\frac{\sqrt{2}}{9} \epsilon_{u} \\
 0 &  -\frac{\sqrt{2}}{9} \epsilon_{u}  &  1
 \end{array} \right) \;\;, \;\; M_{d} \simeq m_{b} \left( \begin{array}{ccc} 
 0 &  d e^{i\omega}  &  0  \\ 
 d e^{-i\omega} &  \frac{2}{9} \epsilon_{d}  &  -\frac{\sqrt{2}}{9} \epsilon_{d} \\
 0 &  -\frac{\sqrt{2}}{9} \epsilon_{d}  &  1 
 \end{array} \right) \] \normalsize where the symmetry breaking parameters $u 
\ll \epsilon_{u} \ll 1$ and $d \ll \epsilon_{d} \ll 1$ are to be determined from 
the known values of quark masses. Using the expressions of Eq.~(\ref{check}), 
one finds $\epsilon_{u} \simeq (9/2) \xi_{ct} \lambda^{4},\, \epsilon_{d} \simeq 
(9/2) \xi_{sb} \lambda^{2},\, u \simeq \sqrt{\xi_{ut}\xi_{ct}} \lambda^{11/2},\, 
d \simeq \sqrt{\xi_{db}\xi_{sb}} \lambda^{3}$ and furthermore the following 
relations which can be regarded as the CKM ``predictions'' of the pattern: 
\begin{eqnarray*} 
\lambda & \simeq & \sqrt{m_{d}/m_{s}} \pm \cos\omega \sqrt{m_{u}/m_{c}} \;, \\
A \lambda^{2} & \simeq & (m_{s}/m_{b}-m_{c}/m_{t})/\sqrt{2} \;, \\
\sigma A \lambda^{3} & \simeq & (m_{s}/m_{b}-m_{c}/m_{t}) \sqrt{m_{u}/2m_{c}} 
\end{eqnarray*} and $\delta \simeq \omega + O(\lambda)$. One sees, when refering 
to the data in Table~\ref{LED}, that this pattern leads to extremely low values 
for $|V_{cb}|$ and $|V_{ub}|$, although it has an acceptable value for the 
quantity $|V_{ub}/V_{cb}|$.  
\smallskip

\begin{flushleft}
(ii) {\it Mass Patterns with Most Texture-zeros}
\end{flushleft}

Starting with Eqs.~(\ref{GP}$-$\ref{constraint_2}), arranging for patterns with 
multiple texture-zeros can be particularly efficient. As an exercise, we insert 
zeros in all possible entries of the mass matrices of Eq.~(\ref{GP}). We find, 
in this way, a total of five allowable five-texture-zero, low energy (at the 
scale of $m_{t}$) patterns. To ensure consistence with the LED, the matrix 
elements of these five-texture-zero patterns are specified in accordance with 
Eq.~(\ref{constraint_1}).  In Table~\ref{max0s} we list these patterns and their 
CKM constraints or ``predictions''~\footnote{``Predictions'' ensue whenever 
certain matrix elements or parameters are overspecified~\cite{MM}. For each of 
the mass patterns in Table~\ref{max0s}, for example, Eq.~(\ref{constraint_1}) 
renders two of the LED parameters dependent (chosen here to be $\lambda$ and 
$\Lambda$ or, equivalently, $|V_{us}|$ and $|V_{ub}|$) while the remaining ones 
are not overspecified and as a result, their experimental values can always be 
accomodated.}.

\scriptsize
\begin{table}
\begin{center}
\begin{tabular}{|c|c|c|c|}
\hline & & & \\
 $\phantom{P}$ & $\tilde{M}_{u}$ & $\tilde{M}_{d}$ & ``Prediction'' \\ & & & 
\\ \hline & & & \\
1 &
$\left( \begin{array}{ccc} 
 0 &  u_{12}\lambda^{6}  &  0  \\   
 u_{12}\lambda^{6} &   u_{22}\lambda^{4}  &  0 \\
 0 &  0  &  u_{33} 
 \end{array} \right)$ & $\left( \begin{array}{ccc} 
 0 &  d_{12}\lambda^{3}  &  0  \\ 
 d^{*}_{12}\lambda^{3} &  d_{22}\lambda^{2}  &  d_{23}\lambda^{2} \\
 0 &  d_{23}\lambda^{2}  &  d_{33} 
 \end{array} \right)$  & $\begin{array}{l}
|V_{us}| = \sqrt{m_{d}/m_{s}} \pm \Delta_{11} \\
\phantom{|V_{cs}| =} + O(\lambda^3) \\ \\
|V_{ub}| = |V_{cb}| \sqrt{m_{u}/m_{c}} \\   
\phantom{|V_{ub}|=} \pm \Delta_{12} + O(\lambda^6)   
\end{array}$ \\ & & & \\ \hline & & & \\
2 &
$\left( \begin{array}{ccc} 
 0 &  u_{12}\lambda^{6}  &  0 \\   
 u_{12}\lambda^{6} &  0  &  u_{23}\lambda^{2} \\
 0 &  u_{23}\lambda^{2}  &  u_{33} 
 \end{array} \right)$ & $\left( \begin{array}{ccc} 
 0 &  d_{12}\lambda^{3}  &  0  \\ 
 d^{*}_{12}\lambda^{3} &  d_{22}\lambda^{2}  &  d_{23}\lambda^{2} \\
 0 &  d^{*}_{23}\lambda^{2}  &  d_{33} 
 \end{array} \right)$  & $\begin{array}{l}
|V_{us}| = \sqrt{m_{d}/m_{s}} \pm \Delta_{21} \\
\phantom{|V_{cs}| =} + O(\lambda^3) \\ \\
|V_{ub}| = |V_{cb}| \sqrt{m_{u}/m_{c}} \\
\phantom{|V_{cs}| =} \pm \Delta_{22} + O(\lambda^6)   
\end{array}$ \\ & & & \\ \hline & & &  \\
3 &
$\left( \begin{array}{ccc} 
 0 &  0  &  u_{13}\lambda^{4}  \\   
 0 &  u_{22}\lambda^{4}  &  0 \\
 u_{13}\lambda^{4} &  0  &  u_{33} 
 \end{array} \right)$ & $\left( \begin{array}{ccc} 
 0 &  d_{12}\lambda^{3}  &  0  \\ 
 d^{*}_{12}\lambda^{3} &  d_{22}\lambda^{2}  &  d_{23}\lambda^{2} \\
 0 &  d_{23}\lambda^{2}  &  d_{33} 
 \end{array} \right)$ & $\begin{array}{l}
|V_{us}| = \sqrt{m_{d}/m_{s}} + \Delta_{31} \\
\phantom{|V_{cs}| =} + O(\lambda^3) \\ \\
|V_{ub}| = \sqrt{m_{u}/m_{t}} \\
\phantom{|V_{cs}| =} \pm \Delta_{32} + O(\lambda^6) 
\end{array}$ \\ & & & \\ \hline & & & \\
4 &
$\left( \begin{array}{ccc} 
 0 &  u_{12}\lambda^{6}  &  0  \\   
 u_{12}\lambda^{6} &  u_{22}\lambda^{4}  &  u_{23}\lambda^{2} \\
 0 &  u_{23}\lambda^{2}  &  u_{33} 
 \end{array} \right)$ & $\left( \begin{array}{ccc} 
 0 &  d_{12}\lambda^{3}  &  0  \\ 
 d^{*}_{12}\lambda^{3} &  d_{22}\lambda^{2}  &  0 \\
 0 &  0  &  d_{33} 
 \end{array} \right)$  & $\begin{array}{l}
|V_{us}| = \sqrt{m_{d}/m_{s}} \pm \Delta_{41} \\
\phantom{|V_{cs}| =} + O(\lambda^3) \\ \\
|V_{ub}| = |V_{cb}| \sqrt{m_{u}/m_{c}} \\
\phantom{|V_{cs}| =}  + \Delta_{42} + O(\lambda^6) 
\end{array}$ \\ & & & \\ \hline & & & \\
5 & 
$\left( \begin{array}{ccc} 
 0 &  0  &  u_{13}\lambda^{4}  \\   
 0 &  u_{22}\lambda^{4}  &  u_{23}\lambda^{2} \\
 u_{13}\lambda^{4} &  u_{23}\lambda^{2}  &  u_{33} 
 \end{array} \right)$ & $\left( \begin{array}{ccc} 
 0 &  d_{12}\lambda^{3}  &  0  \\ 
 d^{*}_{12}\lambda^{3} &  d_{22}\lambda^{2}  &  0 \\
 0 &  0  &  d_{33} 
 \end{array} \right)$  & $\begin{array}{l}
|V_{us}| = \sqrt{m_{d}/m_{s}} \pm \Delta_{51} \\
\phantom{|V_{cs}| =} + O(\lambda^3) \\ \\
|V_{ub}| = \sqrt{m_{u}/m_{c}} \,\{-|V_{cb}|^{2} \\
\phantom{|V_{cs}| =} + m_{c}/m_{t}\}^{1/2} \\
\phantom{|V_{cs}| =} + \Delta_{52} + O(\lambda^6)
\end{array}$ \\ & & & \\ 
\hline
\end{tabular} 
\caption{Quark mass patterns with five texture-zeros and their ``predictions''. 
(The subleading terms $\Delta$'s in the above expresssions for $|V_{cs}|$ and 
$|V_{cb}|$ are relegated to Table~\ref{max0s2}.)} \label{max0s}
\end{center}
\end{table}
\normalsize

\scriptsize
\begin{table}
\begin{center}
\begin{tabular}{|l|l|}
\hline & \\
$\Delta_{11} = \cos\delta \sqrt{\frac{m_{u}}{m_{c}}}$ & $\Delta_{12} =  
\cos\delta \, \frac{\sqrt{m_{d}m_{s}}}{m_{b}} |V_{cb}| $ \\
& \\ \hline & \\
$\Delta_{21} = \cos\delta \sqrt{\frac{m_{u}}{m_{c}}}$ & $\Delta_{22} =  
\cos\delta \, \frac{\sqrt{m_{d}m_{s}}}{m_{b}} (|V_{cb}| + \sqrt{\frac{m_{c}}{m_{t}}})$ 
$\ddag$ \\
& \\ \hline & \\
$\Delta_{31} = 0$ & $\Delta_{32} =  \cos\delta \, \frac{\sqrt{m_{d}m_{s}}}{m_{b}} 
|V_{cb}|$  \\ 
& \\ \hline & \\
$\Delta_{41} = \cos\delta \sqrt{\frac{m_{u}}{m_{c}}}$ & $\Delta_{42} = 0$ \\
& \\ \hline & \\
$\Delta_{51} = \cos\delta \sqrt{\frac{m_{u}}{m_{c}}}(1+\frac{m_{c}}{m_{t}}
|V_{cb}|^{-2})^{-1/2}$ & $\Delta_{52} = 0$ \\     & \\ \hline 
\end{tabular}
\caption{Expressions for the subleading terms in the last column of Table~\ref{max0s}. 
($\ddag$~For definiteness, we assume for this number the special case $\arg\{d_{23}\}=0$ 
and also $d_{23} > 0$.)}  \label{max0s2}
\end{center}
\end{table}
\normalsize

Numerically, with the signs of the quark masses and those of the $\Delta$ terms 
in Table~\ref{max0s} judiciously chosen, the CKM predictions of these patterns 
can be estimated using the quark mass ratios and the value of $|V_{cb}|$ given 
in Table~\ref{LED}. As an example, in Table~\ref{CKM_nums} we give some results, 
corresponding to a certain possible choice of the aforementioned signs. In 
addition we have also included estimates with the much more stringent constraint of 
Eq.~(\ref{mds}) taken into account.

\tiny
\begin{table}
\begin{center}
\begin{tabular}{|c|c|c|}
\hline & & \\
$\phantom{P}$ & $\mbox{\small $\lambda$}$ &  $\mbox{\small $\sigma$}$ \\ & & \\ 
\hline & & \\
1 &
$\begin{array}{c}
(0.23 \pm 0.05) + (0.06 \pm 0.01) \cos\delta \\
\{(0.23 \pm 0.01) + (0.06 \pm 0.01) \cos\delta \}
\end{array}$ & $\begin{array}{c}
(0.27 \pm 0.05) + (0.03 \pm 0.01) \cos\delta \\
\{(0.27 \pm 0.05) + (0.03 \pm 0.01) \cos\delta\}
\end{array}$ \\ & & \\ \hline  & & \\
2 &
$\begin{array}{c}
(0.23 \pm 0.05) + (0.06 \pm 0.01) \cos\delta \\
\{(0.23 \pm 0.01) + (0.06 \pm 0.01) \cos\delta\}
\end{array}$ & $\begin{array}{c}
(0.27 \pm 0.05) + (0.08 \pm 0.03) \cos\delta \\
\{(0.27 \pm 0.05) + (0.08 \pm 0.03) \cos\delta\}
\end{array}$ \\ & & \\ \hline  & & \\
3 &
$\begin{array}{c}
0.23 \pm 0.05 \\
\{0.23 \pm 0.01\}
\end{array}$ & $\begin{array}{c}
(0.42 \pm 0.07) + (0.03 \pm 0.01) \cos\delta \\
\{(0.42 \pm 0.07) + (0.03 \pm 0.01) \cos\delta\}
\end{array}$ \\ & & \\ \hline  & & \\ 
4 &
$\begin{array}{c}
(0.23 \pm 0.05) + (0.06 \pm 0.01) \cos\delta \\
\{(0.23 \pm 0.01) + (0.06 \pm 0.01) \cos\delta\}
\end{array}$ & $\begin{array}{c}
0.27 \pm 0.05 \\
\{0.27 \pm 0.05\}
\end{array}$ \\ & & \\ \hline  & & \\
5 &
$\begin{array}{c}
(0.23 \pm 0.05) + (0.03 \pm 0.01) \cos\delta \\
\{(0.23 \pm 0.01) + (0.03 \pm 0.01) \cos\delta \}
\end{array}$ & $\begin{array}{c}
0.32 \pm 0.08 \\
\{0.32 \pm 0.08\}
\end{array}$ \\ & & \\
\hline
\end{tabular} 
\caption{Numerical estimates for the CKM ``predictions'' of the five-texture-zero 
patterns. (Numbers in the curly brackets are results incorporating the additional 
constraint of Eq.~(\ref{mds}).)} \label{CKM_nums}
\end{center}
\end{table}
\normalsize

To implement the above five-texture-zeros patterns at the GUT scale in the MSSM,
the only necessary modification required of Tables~\ref{max0s} and~\ref{max0s2}  is 
the insertion of the RG scaling factors ($r_{u}$'s, $r_{d}$'s and $r$'s) in 
front of the quark mass ratios and the parameters $V_{cb}$ and $V_{ub}$, based 
on Eqs.~(\ref{ru}$-$\ref{rr}). Having done so, one sees that the CKM predictions of
 patterns 1, 2 and 4 are unaltered to the leading order in $\lambda$ and 
therefore, these patterns are also viable as SUSY GUT patterns; the same is true 
for patterns 3 and 5 for most values of $\tan\beta$~\footnote{This is consistent 
with the findings of Ref.~\cite{RRR} where these five-texture-zero patterns, 
obtained through a detailed numerical analysis, were first presented.}. However, 
near the end regions of the plot in Fig.~\ref{rrr} (where for example $\tan 
\beta$ is very small), the $|V_{ub}|$ predictions of these patterns can become 
unsound. Incidently, as it was observed in Ref.~\cite{RRR}, the nearest 
conceivable six-texture-zero SUSY GUT pattern corresponds to pattern 2 in 
Table~\ref{max0s} with the parameter $d_{23}$ of $\tilde{M}_{d}$ set to zero. As a 
result, this pattern generates an extra, but unfortunately generally 
unfavorable, CKM ``prediction'' 
$|V_{cb}|=\{\sqrt{r_{u}/r^{2}}\}\sqrt{m_{c}/m_{t}} + O(\lambda^4)$ (since the 
ratio $r_{u}/r^{2}$ is typically close to 1 according to Fig.~\ref{rrr}). 
Nonetheless, in light of our discussion on the evolution of the LED parameters, 
this six-texture-zero pattern could still be viable, should the scenario in 
which $\tan\beta \ll O(m_{t}/m_{b})$ (consequently $r_{u}/r^{2} \simeq r$) and 
furthermore $r \simeq O(\lambda)$ prevails.

The detailed CKM ``predictions'' of the five patterns given here can be used to 
further speculate in favor of (or against) them, especially if experimental data 
becomes more precise. For instance, taking into account Eq.~(\ref{mds}) one sees 
from Tables~\ref{max0s2} and \ref{CKM_nums} that for patterns~1, 2 and 4 to be 
successful, the CP phase $\delta$ must be quite large. The same is true for 
pattern~5, although to a slightly lesser degree. On the other hand, pattern~3 
imposes no such restriction, instead it favors a somewhat larger $|V_{ub}|$ when 
compared to the rest.
\smallskip

\begin{flushleft}
(iii) {\it Mass Patterns Useful as Templates}
\end{flushleft}

By relating the matrix elements in Eq.~(\ref{GP}) which may have similar orders 
of magnitude, one can search or arrange for patterns that have fewer independent 
parameters and thus that have potentially greater predictive power. Below, we 
provide five such simple quark mass patterns in Table~\ref{templates} (and the 
CKM ``predictions'' of these patterns in Table~\ref{templates2}) with the hope 
that they may be useful as templates for contemplating quark mass 
Ansatze.~\footnote{ To implement these patterns at the GUT scale in the MSSM, 
one simply takes into account the RG scaling of the quark mass ratios and the 
CKM parameters in Table~\ref{templates2}, in complete analogy to the previous
case of five-texture-zero patterns.}    

\scriptsize
\begin{table}[t]
\begin{center}
\begin{tabular}{|c|c|c|}
\hline & & \\
 $\phantom{P}$ & $\tilde{M}_{u}$ & $\tilde{M}_{d}$ \\  & & \\ \hline & & \\
1 &
$\left( \begin{array}{ccc} 
 \lsim O(\lambda^{9}) &  \lsim O(\lambda^{6})  &  y_{u}B e^{-i\phi_{u}}\lambda^{4}  
\\    \lsim O(\lambda^{6}) &   B\lambda^{4}  & x_{u}B\lambda^{4} \\
 y_{u}B e^{i\phi_{u}}\lambda^{4} &  x_{u}B\lambda^{4}  &  A 
 \end{array} \right)$ & $\left( \begin{array}{ccc} 
  \lsim O(\lambda^{6}) &  Fe^{-i\psi_{d}}\lambda^{3}  &  y_{d}F e^{-i\phi_{d}}
\lambda^{3}  \\ 
 Fe^{i\psi_{d}}\lambda^{3} &  E\lambda^{2}  &  x_{d}E\lambda^{2} \\
 y_{d}F e^{i\phi_{d}}\lambda^{3} &  x_{d}E\lambda^{2}  &  D 
 \end{array} \right)$ \\ & & \\ 
\hline & & \\
2 &
$\left( \begin{array}{ccc} 
 C\lambda^{7} &  z_{u}C\lambda^{7}  &  y_{u}B e^{-i\phi_{u}}\lambda^{4}  \\   
 z_{u}C\lambda^{7} &   B\lambda^{4}  & x_{u}B\lambda^{4} \\
 y_{u}B e^{i\phi_{u}}\lambda^{4} &  x_{u}B\lambda^{4}  &  A 
 \end{array} \right)$ & $\left( \begin{array}{ccc} 
 \lsim O(\lambda^{6}) &  Fe^{-i\psi_{d}}\lambda^{3}  &  y_{d}F e^{-i\phi_{d}}
\lambda^{3}  \\ 
 Fe^{i\psi_{d}}\lambda^{3} &  E\lambda^{2}  &  x_{d}E\lambda^{2} \\
 y_{d}F e^{i\phi_{d}}\lambda^{3} &  x_{d}E\lambda^{2}  &  D 
 \end{array} \right)$ \\ & & \\ 
\hline & & \\
3 &
$\left( \begin{array}{ccc} 
 \lsim O(\lambda^{9}) &  C\lambda^{6}  &  \phantom{b}\lsim O(\lambda^{6})\phantom{b}  
\\      C\lambda^{6} &   \lsim O(\lambda^{6})  & B\lambda^{2} \\
 \phantom{b}\lsim O(\lambda^{6})\phantom{b} &  B\lambda^{2}  &  A 
 \end{array} \right)$ & $\left( \begin{array}{ccc} 
  \lsim O(\lambda^{6}) &  Fe^{-i\psi_{d}}\lambda^{3}  &  y_{d}F e^{-i\phi_{d}}
\lambda^{3}  \\ 
 Fe^{i\psi_{d}}\lambda^{3} &  E\lambda^{2}  &  x_{d}E\lambda^{2} \\
 y_{d}F e^{i\phi_{d}}\lambda^{3} &  x_{d}E\lambda^{2}  &  D 
 \end{array} \right)$ \\ & & \\ 
\hline & & \\
4 &
$\left( \begin{array}{ccc} 
 C\lambda^{7} &  z_{u}C\lambda^{7}  &  \phantom{b} \lsim O(\lambda^{6}) 
\phantom{b} \\   
 z_{u}C\lambda^{7} &   \lsim O(\lambda^{6})  & B\lambda^{2} \\
  \phantom{b}\lsim O(\lambda^{6}) \phantom{b} &  B\lambda^{2}  &  A 
 \end{array} \right)$ & $\left( \begin{array}{ccc} 
 \lsim O(\lambda^{6}) &  Fe^{-i\psi_{d}}\lambda^{3}  &  y_{d}F    e^{-i\phi_{d}}
\lambda^{3}  \\ 
 Fe^{i\psi_{d}}\lambda^{3} &  E\lambda^{2}  &  x_{d}E\lambda^{2} \\
 y_{d}F e^{i\phi_{d}}\lambda^{3} &  x_{d}E\lambda^{2}  &  D 
 \end{array} \right)$ \\ & & \\
\hline & & \\ 
5 &
$\left( \begin{array}{ccc} 
 \lsim O(\lambda^{9}) &  \lsim O(\lambda^{8}) &  \phantom{BB}C\lambda^{4}\phantom{BB}  
\\     \lsim O(\lambda^{8})  &   x_{u}C\lambda^{4}  & B\lambda^{2} \\
 \phantom{BB}C\lambda^{4}\phantom{BB} & B\lambda^{2}  &  A 
 \end{array} \right)$ & $\left( \begin{array}{ccc} 
  \lsim O(\lambda^{6}) &  Fe^{-i\psi_{d}}\lambda^{3}  &  y_{d}F e^{-i\phi_{d}}
\lambda^{3}  \\ 
 Fe^{i\psi_{d}}\lambda^{3} &  E\lambda^{2}  &  x_{d}E\lambda^{2} \\
 y_{d}F e^{i\phi_{d}}\lambda^{3} &  x_{d}E\lambda^{2}  &  D 
 \end{array} \right)$ \\ & & \\ 
\hline 
\end{tabular} 
\caption{Quark mass patterns which may be useful as templates. (In the above, 
$(A, B ... F)$ are fixed parameters of $O(1)$ and $(x, y, z)$'s are adjustable 
parameters.)} \label{templates}
\end{center}
\end{table}
\normalsize

\scriptsize
\begin{table}
\begin{center}
\begin{tabular}{|c|c|c|c|}
\hline & & & \\
 $\phantom{P}$ & $|V_{us}|$ & $|V_{cb}|$ & $V_{ub}$ \\ & & & \\ \hline & & & \\
1 & $\begin{array}{l}
\phantom{+} \sqrt{\frac{m_{d}}{m_{s}}} \rule{0mm}{4mm} \\ \\
 \pm \cos\psi_{d}\sqrt{\frac{m_{u}}{m_{c}} + y^{2}_{u}\frac{m_{c}}{m_{t}}} 
\rule{0mm}{4mm} \\ \\
+ O(\lambda^3) \rule{0mm}{4mm}
\end{array}$ &
$\begin{array}{l}
\phantom{+} x_{d}\frac{m_{s}}{m_{b}} \rule{0mm}{4mm}\\ \\
+ x_{u}\frac{m_{c}}{m_{t}} \rule{0mm}{4mm} \\ \\
+ O(\lambda^4) \rule{0mm}{4mm}
\end{array}$ &
$\begin{array}{l}
\left\{ \rule{0mm}{3.5mm} y_{d}\frac{\sqrt{m_{d}m_{s}}}{m_{b}} e^{-i\phi_{d}}  + 
y_{u}\frac{m_{c}}{m_{t}} e^{-i\phi_{u}} \right. \\ \\ 
\left. \pm \sqrt{\frac{m_{u}}{m_{c}} + y^{2}_{u}\frac{m_{c}}{m_{t}}}\, |V_{cb}|  
\rule{0mm}{3.5mm} \right\} e^{i\psi_{d}} \\ \\
+ x_{d}(\frac{m_{s}}{m_{b}})^{2}|V_{us}| + O(\lambda^6) 
\end{array}$ \\ & & & \\ 
\hline & & & \\
2 & $\begin{array}{l}
\phantom{+} \sqrt{\frac{m_{d}}{m_{s}}} \rule{0mm}{4mm} \\ \\
+ z_{u}(\frac{m_{u}}{m_{c}} + y^{2}_{u}\frac{m_{c}}{m_{t}})\cos\psi_{d} 
\rule{0mm}{4mm} \\ \\
+ O(\lambda^4) \rule{0mm}{4mm}
\end{array}$ &
$\begin{array}{l}
\phantom{+} x_{d}\frac{m_{s}}{m_{b}} \rule{0mm}{4mm} \\ \\
+ x_{u}\frac{m_{c}}{m_{t}} \rule{0mm}{4mm} \\ \\
+ O(\lambda^4) \rule{0mm}{4mm}
\end{array}$ &
$\begin{array}{l}
\left\{ \rule{0mm}{3.5mm} y_{d}\frac{\sqrt{m_{d}m_{s}}}{m_{b}} e^{-i\phi_{d}} + 
y_{u}\frac{m_{c}}{m_{t}} e^{-i\phi_{u}} \right. \\ \\ 
\left. + z_{u}(\frac{m_{u}}{m_{c}} + y^{2}_{u}\frac{m_{c}}{m_{t}}) |V_{cb}| 
\rule{0mm}{3.5mm} \right\} e^{i\psi_{d}} \\ \\
+ x_{d}(\frac{m_{s}}{m_{b}})^{2}|V_{us}| + O(\lambda^6)
\end{array}$ \\ & & & \\ 
\hline & & & \\
3 & $\begin{array}{l}
\phantom{+} \sqrt{\frac{m_{d}}{m_{s}}} \rule{0mm}{4mm} \\ \\
 \pm \cos\psi_{d}\sqrt{\frac{m_{u}}{m_{c}}}\phantom{+ y^{2}_{u}\frac{m_{c}}{m_{t}}} 
\rule{0mm}{4mm} \\ \\
+ O(\lambda^3) \rule{0mm}{4mm}
\end{array}$ &
$\begin{array}{l}
\phantom{+} \sqrt{\frac{m_{c}}{m_{t}}} \rule{0mm}{4mm} \\ \\
- x_{d}\frac{m_{s}}{m_{b}} \rule{0mm}{4mm} \\ \\
+ O(\lambda^4) \rule{0mm}{4mm}
\end{array}$ &
$\begin{array}{l}
\left\{ \rule{0mm}{3.5mm} y_{d}\frac{\sqrt{m_{d}m_{s}}}{m_{b}} e^{-i\phi_{d}} \phantom{y_{u}\frac{m_{c}}{m_{t}} e^{-i\phi_{u}}} \right. \\ \\ 
\left. \pm  \sqrt{\frac{m_{u}}{m_{c}}} |V_{cb}| \rule{0mm}{3.5mm} \right\} 
e^{i\psi_{d}} \\ \\
+ x_{d}(\frac{m_{s}}{m_{b}})^{2}|V_{us}| + O(\lambda^6)
\end{array}$ \\ & & & \\ 
\hline & & & \\
4 & $\begin{array}{l}
\phantom{+} \sqrt{\frac{m_{d}}{m_{s}}} \rule{0mm}{4mm} \\ \\
+ z_{u}\frac{m_{u}}{m_{c}} \cos\psi_{d} \phantom{+ y^{2}_{u}\frac{m_{c}}{m_{t}}} 
\rule{0mm}{4mm} \\ \\
+ O(\lambda^4) \rule{0mm}{4mm}
\end{array}$ &
$\begin{array}{l}
\phantom{+} \sqrt{\frac{m_{c}}{m_{t}}} \rule{0mm}{4mm}\\ \\
- x_{d}\frac{m_{s}}{m_{b}} \rule{0mm}{4mm} \\ \\
+ O(\lambda^4) \rule{0mm}{4mm}
\end{array}$ &
$\begin{array}{l}
\left\{ \rule{0mm}{3.5mm} y_{d}\frac{\sqrt{m_{d}m_{s}}}{m_{b}} e^{-i\phi_{d}} \phantom{y_{u}\frac{m_{c}}{m_{t}} e^{-i\phi_{u}}} \right. \\ \\ 
\left. + z_{u} \frac{m_{u}}{m_{c}} |V_{cb}| \rule{0mm}{3.5mm} \right\} 
e^{i\psi_{d}} \\ \\
+ x_{d}(\frac{m_{s}}{m_{b}})^{2}|V_{us}| + O(\lambda^6)
\end{array}$ \\ & & & \\
\hline & & & \\
5 & $\begin{array}{l}
\phantom{+} \sqrt{\frac{m_{d}}{m_{s}}} \rule{0mm}{4mm} \\ \\
+ \cos\psi_{d} \sqrt{\frac{m_{u}}{m_{c}}(1+w)} \rule{0mm}{4mm} \\ \\
+ O(\lambda^4) \rule{0mm}{4mm} \; $\ddag$
\end{array}$ &
$\begin{array}{l}
\phantom{+} \sqrt{\frac{m_{c}}{m_{t}}(1+w^{-1})} \rule{0mm}{4mm} \\ \\
- x_{d}\frac{m_{s}}{m_{b}} \rule{0mm}{4mm} \\ \\
+ O(\lambda^4) \rule{0mm}{4mm} \; $\ddag$
\end{array}$ &
$\begin{array}{l}
\left\{ \rule{0mm}{3.5mm} y_{d}\frac{\sqrt{m_{d}m_{s}}}{m_{b}} e^{-i\phi_{d}} - 
\sqrt{\frac{m_{u}}{m_{t}} w} \right. \\ \\ 
 +  \left. \sqrt{\frac{m_{u}}{m_{c}}(1+w)}\, |V_{cb}| \rule{0mm}{3.5mm} \right\} 
e^{i\psi_{d}} \\ \\
+ x_{d}(\frac{m_{s}}{m_{b}})^{2}|V_{us}| + O(\lambda^6)  \; $\ddag$
\end{array}$ \\ & & & \\
\hline
\end{tabular} 
\caption{CKM ``Predictions'' of the patterns in Table~\ref{templates}.($\ddag$ In 
these expressions, $w$ is defined to be the quantity $\left\{\frac{m_{c}^{2}}{x^{2}_{u}m_{u}m_{t}}\right\}^{\frac{1}{3}}$ for 
notational brevity.)} \label{templates2}
\end{center}
\end{table}
\normalsize

In deriving the results of Tables~\ref{templates} and~\ref{templates2}, the 
parameters $(x, y, z)$'s are assumed to be of $O(1)$ or less, but are otherwise 
unspecified. This allows for certain flexibility in pattern-fitting. Evidently, 
not all values for the $(x, y, z)$'s work equally well; using 
Table~\ref{templates2} and the LED of Table~\ref{LED} however, one can readily 
determine the feasibility of a given set of values for these adjustable 
parameters.    

Certainly, construction of more elaborate patterns is also possible. But 
already, a host of interesting patterns can be obtained from 
Table~\ref{templates}. In particular, notice that texture-zeros can easily be 
accomodated by inserting them where allowed or by selectively specifying some of 
the $(x, y, z)$'s to be 0's.~\footnote{In fact, the five-texture-zero patterns 
of Table~\ref{max0s} can be gotten this way as well.} Similarly, equalities 
among matrix elements can be arranged by specifying some of the $(x, y, z)$'s to 
be 1's. As an illustrative example, let us choose in pattern 2, $y_{d}=0 \;,\; 
x_{u}=y_{u}=z_{u}=x_{d}=1$, and $\psi_{d}=\pi$; the result is a rather simple 
looking pattern in which \small \[\tilde{M}_{u} = \left( \begin{array}{ccc} 
 C\lambda^{7} &  C\lambda^{7} &  B\lambda^{4}e^{-i\phi_{u}}  \\   
 C\lambda^{7} &   B\lambda^{4}  & B\lambda^{4} \\
  B\lambda^{4}e^{i\phi_{u}} &  B\lambda^{4}  &  A 
 \end{array} \right) \;, \quad \tilde{M}_{d}=\left( \begin{array}{ccc} 
 0 &  F\lambda^{3}  &  0 \\ 
 F\lambda^{3} &  E\lambda^{2}  &  E\lambda^{2} \\
 0 &  E\lambda^{2}  &  D 
 \end{array} \right)\;. \] \normalsize As a low energy pattern, it gives, to a 
very good approximation, the CKM ``predictions'': \small \[|V_{us}| \simeq 
\sqrt{\frac{m_{d}}{m_{s}}} - \frac{m_{u}}{m_{c}} - \frac{m_{c}}{m_{t}},\; 
|V_{cb}| \simeq \frac{m_{s}}{m_{b}} + \frac{m_{c}}{m_{t}},\; |V_{ub}| \simeq 
\frac{m_{c}}{m_{t}} \] \normalsize and $\delta \simeq \phi_{u}-\pi$ (which of 
course yields the correct value for $\delta$ automatically when $\phi_{u}$ is 
suitably chosen). To check the soundness of the above ``predictions'', we input 
the quark mass ratios of Table~\ref{LED} and  find \[ |V_{us}| \simeq 0.22 \pm 
0.05 \;,\; |V_{cb}| \simeq 0.030 \pm 0.009\;,\; \mbox{and} \; |V_{ub}| \simeq 
0.0034 \pm 0.0003 \] in reasonable agreement with the CKM data, also given in 
Table~\ref{LED}. (If we incorporate Eq.~(\ref{mds}) into the calculation of 
$|V_{us}|$ above, we have instead $|V_{us}| \simeq 0.22 \pm 0.01$.) 
Correspondingly, as a SUSY GUT pattern, it predicts: \small \[|V_{us}| \simeq 
\sqrt{\frac{m_{d}}{m_{s}}} - \frac{m_{u}}{m_{c}} - 
\left\{r_{u}\right\}\frac{m_{c}}{m_{t}},\, |V_{cb}| \simeq 
\left\{\frac{r_{d}}{r}\right\}\frac{m_{s}}{m_{b}} + 
\left\{\frac{r_{u}}{r}\right\}\frac{m_{c}}{m_{t}},\, |V_{ub}| \simeq 
\left\{\frac{r_{u}}{r}\right\}\frac{m_{c}}{m_{t}} \] \normalsize and again 
$\delta \simeq \phi_{u}-\pi$. Since for most values of $\tan\beta$, $r_{d}/r 
\simeq 1$ and $r_{u}/r$ is of $O(1)$ (Fig.~\ref{rrr}), one sees that these GUT 
pattern ``predictions'' are equally ``sound'', with possible exceptions noted 
for extreme values of $\tan\beta$. 
\bigskip

It is worth pointing out that using the results of 
Eqs.~(\ref{GP}$-$\ref{constraint_2}), the quark mass patterns of our examples 
are constructed in a completely systematic and often very efficient manner; 
moreover, ``predictions'' of these mass patterns are readily obtained in the 
form of explicit analytical expressions -- making transparent the viability 
conditions of each pattern.      
\bigskip

\begin{flushleft}
\bf{Acknowledgements}
\end{flushleft} 
\rm

I would like to thank Professor Roberto Peccei for suggesting this project and 
for many valuable comments. This work is supported in part by the Department of 
Energy under Grant No. FG03-91ER40662.

\end{document}